# Near-infrared fluorescent nanoprobes for irreversibility in nonequilibrium actomyosin networks

Adi Hendler-Neumark[a], Itamar Magar[b], Shirel Kleiner[a], Geffen Rosenberg[a], and Gili Bisker[a,c,d,e,f]*

[a] School of Biomedical Engineering, Faculty of Engineering, Tel Aviv University, Tel Aviv 6997801, Israel

[b] School of Physics and Astronomy, Faculty of Exact Sciences, Tel Aviv University, Tel Aviv 6997801, Israel

[c] Center for Physics and Chemistry of Living Systems, Tel Aviv University, Tel Aviv 6997801, Israel

[d] Center for Nanoscience and Nanotechnology, Tel Aviv University, Tel Aviv 6997801, Israel

[e] Center for Light Matter Interaction, Tel Aviv University, Tel Aviv 6997801, Israel

[f] Sagol School of Neuroscience, Tel Aviv University, Tel Aviv 6997801, Israel

*E-mail: bisker@tauex.tau.ac.il

**Abstract**

Actomyosin networks operate far from equilibrium, yet detecting the onset of motor-driven irreversible dynamics remains challenging. Here, we embed near-infrared (NIR) fluorescent single-walled carbon nanotubes (SWCNTs) within reconstituted actin networks, and use their nonphotobleaching emission to optically report ATP-powered myosin contractile activity. G-actin-dispersed SWCNTs are incorporated into polymerized F-actin without perturbing network assembly, enabling long-term, single-emitter fluorescence monitoring. Upon myosin addition, the NIR fluorescence levels of individual SWCNTs exhibit enhanced temporal fluctuations, and population-level statistics reveal deviations from equilibrium behaviour. The index of dispersion (IOD) distributions shift and broaden relative to equilibrium baselines, and the Kullback-Leibler divergence between IOD distributions systematically increases with increasing motor activity. Stationarity analysis further shows a dose-dependent increase in the fraction of nonstationary fluorescence traces, indicating the emergence of irreversible, time-evolving dynamics. These results establish SWCNTs as minimally invasive optical probes of irreversibility in nonequilibrium actomyosin assemblies, with broad applicability to other active biopolymer systems.

## **Main**

Stochastic systems maintained far from thermodynamic equilibrium are ubiquitous in nature[1,2]. Biological systems, including cells, tissues, and whole organisms, inherently operate under nonequilibrium conditions, continuously consuming energy to drive mechanical work, biochemical turnover, and information processing[3–5]. Cells function as dissipative dynamical systems that exchange energy, matter, and information with their environment[2], through networks of biochemical reactions[6–9], sustained by the utilization of nutrient supply and the hydrolysis of chemical fuel molecules such as adenosine triphosphate (ATP)[10–12]. As such, cellular processes, including cell division, migration, and replication, are fundamentally nonstationary and irreversible, with time-asymmetric fluctuations[3], and require dynamic reorganization and remodelling of the cytoskeleton and other cellular components[3,13–16]. Detecting and quantifying signatures of irreversibility in the context of nonequilibrium dynamics, which is directly linked to dissipation and the production of entropy[5,17–23], is therefore essential for understanding the physical principles that underlie cellular behaviour and living systems[3].

A paradigmatic example of a biological system that operates out of equilibrium is the actomyosin network, a key component of the cytoskeleton, responsible for generating contractile forces in both muscle and non-muscle cells[24,25]. This composite system consists primarily of filamentous actin (F-actin), formed by the polymerization of globular actin (G-actin) monomers into long helical filaments[26], and myosin motor proteins, ATPases that convert the free energy of ATP hydrolysis into directed mechanical motion along actin filaments. By coupling ATP consumption to filament sliding, myosin produces active stresses that remodel the network and drive contractile forces[13,15,16].

Reconstituted actomyosin assemblies have emerged as a powerful minimal model system for studying how microscopic motor-generated activity regulates macroscopic mechanical properties and fluctuations under nonequilibrium conditions[3,27–32]. Such systems have revealed that active stresses driven by myosin motors can induce dynamic reorganization, enhanced fluctuations, filament buckling, mechanical stiffening, network-level contraction, and generally, the emergent nonequilibrium behaviour[3,14]. Experimental approaches to probe these dynamics typically rely on microrheology or fluorescently labelled filaments[3,33–40]. However, tracking the position of micro-sized beads used in microrheology setups has limited spatial resolution, while conventional fluorescent dyes suffer from photobleaching and blinking under prolonged illumination[41].

Single-walled carbon nanotubes (SWCNTs) have been broadly established as reporters of dynamical processes in soft and biological matter[42]. These nanoscale cylinders, conceptualized as rolled graphene sheets with defined chiralities[43], exhibit intrinsic and exceptionally photostable near-infrared (NIR) fluorescence[44,45], which overlaps with the biologically transparent spectral window, where scattering, absorption, and autofluorescence are minimal, thereby offering a higher signal-to-noise ratio, compared to the visible range[46–50]. Prior work has leveraged these properties to track nanoscale rearrangements in viscoelastic supramolecular gels[33,51], map nanoscale morphology and inner diffusivity in the synaptic extracellular space[52], and quantify nonequilibrium activity in reconstituted systems through the position[53] or bending-mode dynamics[54–56] of individual nanotubes.

Another important optical property of SWCNTs is the sensitivity of their NIR fluorescence to the local microenvironment[57–59]. Changes in the dielectric surroundings, nanoscale strain, adsorption/desorption events, and local molecular rearrangements at their surface can all modulate SWCNT fluorescence intensity[57]. Critically, unlike microrheology or other experimental settings that track particle positions or nanotube shapes, SWCNTs can report on fluctuations via their emission intensity, enabling continuous, non-invasive, single-molecule readouts of local activity with high temporal resolution[59–64]. Therefore, they offer a promising alternative as optical probes for microscopic nonequilibrium systems.

In this work, we employ SWCNTs as embedded NIR fluorescent reporters to detect the onset of irreversible, nonstationary dynamics in reconstituted actin networks activated by ATP-fuelled myosin molecular motors. SWCNTs are first suspended by G-actin and subsequently incorporated into polymerizing F-actin without perturbing network assembly, enabling long-term, single-emitter fluorescence measurements. Comparing the SWCNT NIR emission with that of a conventional fluorescent dye used to label actin filaments, we demonstrate the superior photostability and suitability of SWCNT for extended imaging. Myosin motors are then introduced at varying concentrations to drive the networks out of equilibrium through the contractile activity induced by ATP hydrolysis. We analyze the resulting NIR fluorescence time series before and after myosin addition, compute the index of dispersion (IOD) in sliding windows of different sizes, compare the corresponding IOD distributions to pre-intervention baselines using the Kullback-Leibler divergence (KLD), and perform stationarity tests on the temporal fluorescence trajectories. These complementary statistical descriptors reveal a robust and reproducible shift from stationary baseline behaviour to a nonstationary, irreversible regime following activation by molecular motors, demonstrating that SWCNT fluorescence sensitively reports the onset of nonequilibrium dynamics in the actomyosin network. Importantly, our approach does not rely on tracking particle positions or filament shapes, which are constrained by optical resolution, but instead extracts dynamical information solely from fluorescence intensity fluctuations that are sensitive to nanoscale changes in the SWCNTs' local environment. Together, these results establish SWCNTs as robust nanoscale probes for nonequilibrium behaviour in cytoskeletal systems, with potential applicability to other active biopolymer and soft-matter systems.

## Incorporation of G-actin-SWCNTs into the F-actin network

Actomyosin is a fundamental component of the cytoskeletal contractile machinery within cells. These assemblies consist of F-actin, formed by the polymerization of G-actin monomers, and myosin motors, which hydrolyse ATP to generate mechanical force and drive sliding motion along the filaments[13]. To incorporate NIR fluorescent probes into an actin network driven by myosin molecular motors, we first established an actin-mediated strategy for dispersing and integrating SWCNTs into the actin network. SWCNTs were directly dispersed using purified monomeric G-actin, enabling their subsequent incorporation into the polymerized F-actin network. The SWCNTs were first tip-sonicated with G-actin, producing a stable colloidal suspension (G-actin-SWCNTs) as confirmed by the sharp, distinct absorption peaks corresponding to different SWCNT chiralities (**Figure 1**A).

High-resolution atomic force microscopy (AFM) imaging revealed the coexistence of G-actin monomers and G-actin-SWCNTs within the suspension (**Figure 1**B). The measured height of the nanotubes was approximately 1-2 nm (**Figure 1**B, **i**), consistent with the dimensions of individual SWCNTs. Additional features could be attributed to actin monomers (~7 nm, **Figure 1**B, **ii**), actin aggregates or nucleations of 20-30 nm in diameter[63] (**Figure 1**B, **iii**), and SWCNTs functionalized by G-actin (~10 nm, **Figure 1**B, **iv**). Notably, features of a few nanometers up to ~30 nm in height, attributed to the actin monomers and the small actin aggregates, were likewise present in the AFM image of the actin-only control (**Figure S1**A), indicating that SWCNT addition did not alter actin morphology. Complementary transmission electron microscopy (TEM) of the G-actin-SWCNTs sample (**Figure 1**C) revealed elongated tubular structures consistent with nanotubes, and G-actin bundles that were also observed in the actin-only controls (**Figure S1**B). The NIR emission of the G-actin-SWCNT suspension exhibited the expected chirality-resolved peaks characteristic of the optical properties of a well-dispersed SWCNT suspension (**Figure 1**D).

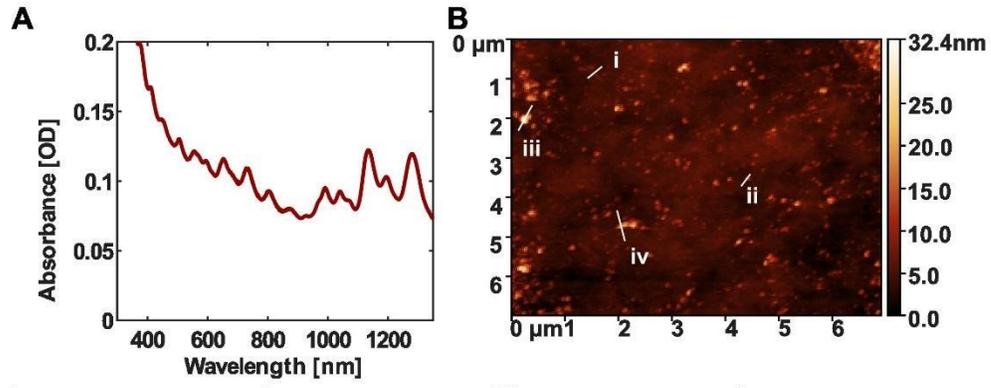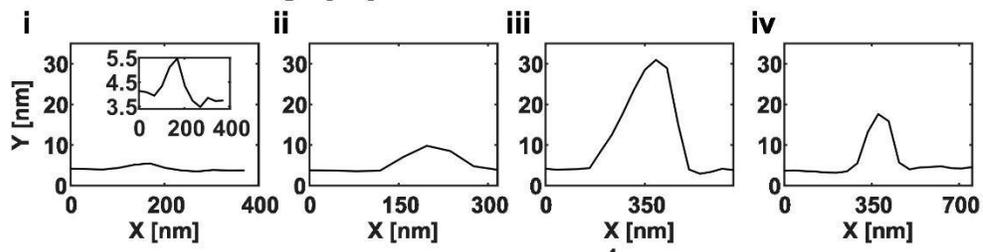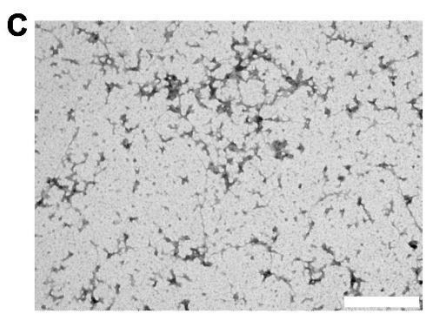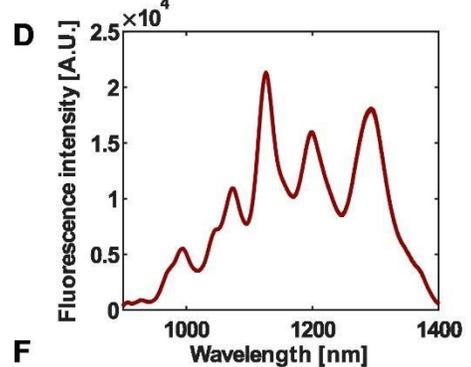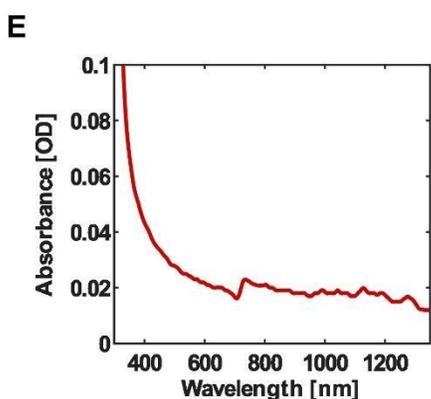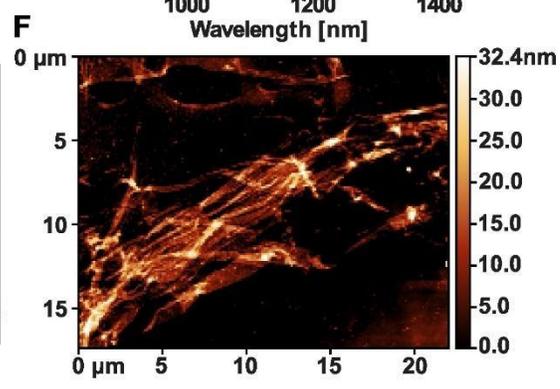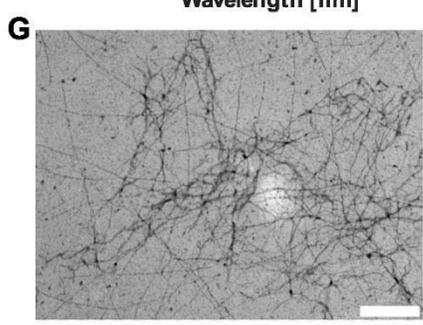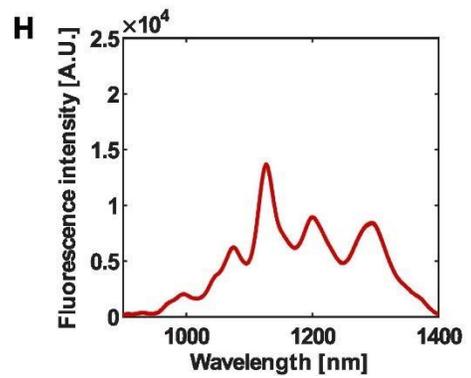

**Figure 1: Characterization of G-actin-SWCNTs and F-actin-SWCNTs.** A) Absorption spectrum of G-actin-SWCNTs. B) AFM images of G-actin-SWCNTs. The marked lines represent: i. SWCNT, ii. G-actin monomer, iii. G-actin aggregate, iv. G-actin-SWCNT, along with the corresponding height profile. C) TEM images of G-actin-SWCNTs. Scale bar is 500 nm. D) NIR fluorescence spectrum of G-actin-SWCNTs under 730 nm excitation. E) Absorption spectrum of F-actin-SWCNTs. F) AFM images of F-actin-SWCNTs. G) TEM images of F-actin-SWCNTs. Scale bar is 500 nm. H) NIR fluorescence spectrum of F-actin-SWCNTs under 730 nm excitation.

G-actin-SWCNTs were next introduced into a standard F-actin polymerization mixture containing G-actin and ATP to form F-actin-SWCNT composites. The colloidal stability of the F-actin-SWCNTs within the network was confirmed by the characteristic chirality-resolved absorption peaks (**Figure 1**E). AFM and TEM images of the resulting F-actin-SWCNT samples (**Figure 1**F,G) revealed an extended filamentous network, with a similar morphology as that of control F-actin samples without SWCNTs (**Figure S1**C,D), indicating that the SWCNT loading did not perturb the actin polymerization. The fluorescence spectrum of the F-actin-SWCNT sample under 730 nm excitation displayed the expected chirality-dependent NIR emission (**Figure 1**H), qualitatively matching the emission peaks of G-actin-SWCNTs (**Figure 1**D), but with reduced intensity. This intensity decrease reflects the deliberate dilution of the G-actin-SWCNT stock in the polymerization mixture during network formation to achieve a sparse, well-separated SWCNT distribution, minimizing background, enabling robust NIR-image images, and, critically, avoiding perturbation of actin polymerization and myosin/ATP-driven function.

### Integration of non-photobleaching G-actin–SWCNTs into the F-actin network

Having formed the composite system of the G-actin-SWCNTs incorporated within the F-actin network, the deliberately low concentration of SWCNTs, chosen to avoid perturbing actin polymerization, was also helpful in preventing NIR overexposure and in capturing signal from spatially separated SWCNTs. To localize SWCNTs within the filamentous network, we performed co-registered dual-channel imaging of phalloidin-labelled F-actin in the visible (490 nm LED excitation and 505-530 nm emission) and SWCNT fluorescence in the NIR (730 nm CW laser excitation and >900 nm emission), acquired as z-stacks through the three-dimensional sample to assess incorporation. Visible fluorescence was recorded using an EMCCD camera (**Figure 2**A), while NIR fluorescence was acquired with an InGaAs camera (**Figure 2**B). The two channels were then registered and overlaid (**Figure 2**C), revealing that both fluorescence signals came into focus within the same z-plane, confirming that the SWCNTs were successfully incorporated into the F-actin network.

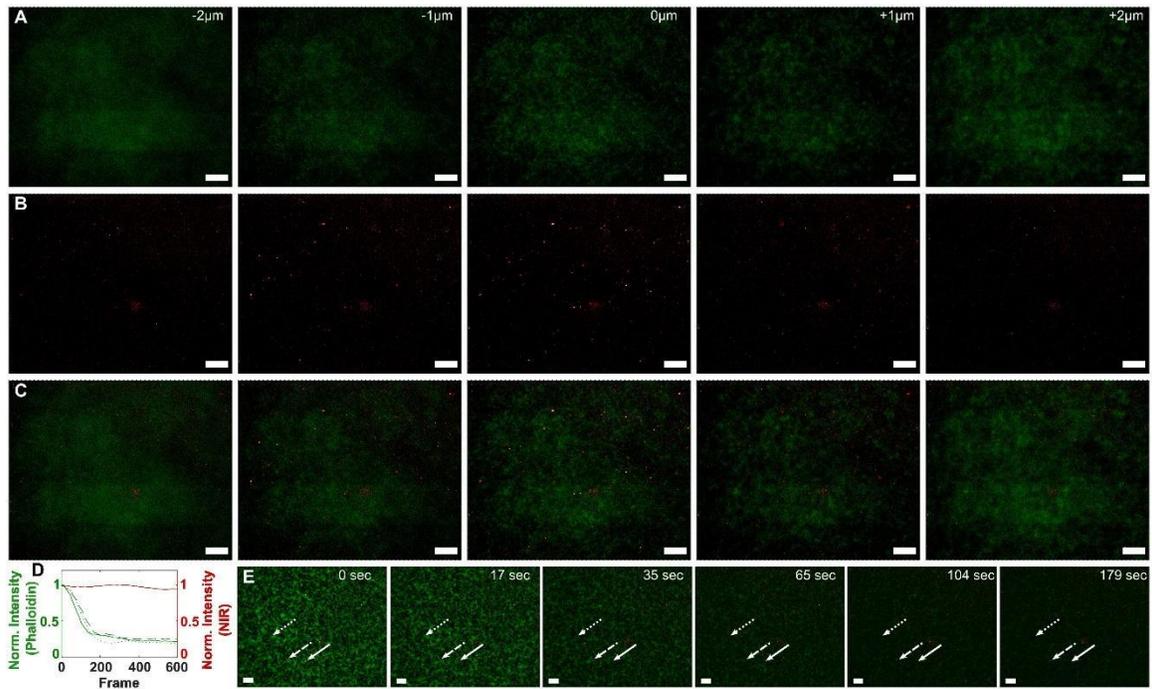

**Figure 2: Integration and co-localization of G-actin-SWCNT within F-actin network.** A) Z-stack images of the phalloidin-stained F-actin-SWCNT network in the visible channel (Acti-stain 488, $\lambda_{ex}$ = 490 nm, $\lambda_{em}$ = 505-530 nm), in the B) NIR channel (SWCNTs, $\lambda_{ex}$ = 730 nm, $\lambda_{em}$ > 900 nm), and C) overlaid channels. Columns correspond to vertical positions of -2 μm, -1 μm, 0 μm, +1 μm, and 2μm, relative to the focal plane. Scale bars are 10 μm. D) Normalized intensity of three representative ROIs of the Acti-stain 488 phalloidin visible fluorescence (solid, dashed, and dotted green lines) and of the SWCNT NIR fluorescence (red curve) over the entire FOV, during a 600-frame time-lapse movie (300 ms intervals). E) Overlay snapshots adapted from **Movie S1** of the F-actin-SWCNTs in the two fluorescent channels, Acti-stain 488 phalloidin (green) and SWCNTs (red). Arrows

point to ROIs of the visible fluorescence presented in panel D (green lines). Scale bars are 10 μm.

Imaging the F-actin network with the phalloidin dye posed a challenge due to photobleaching, a common issue typical of organic fluorophores under repeated excitation[41]. In contrast, SWCNT fluorescence emission is known for its exceptional photostability, as it neither photobleaches nor blinks[59,64,65]. In dual-channel time-lapse imaging, the NIR SWCNT signal remained essentially constant, while the fluorescence of the Acti-stain 488 phalloidin dye decreased rapidly over time, reaching approximately 20% of the original intensity after ~200 frames (**Figure 2**D and **movie S1**). Initially, the actin network was clearly visible owing to the dye, but within ~35 seconds, a loss of visible contrast was evident (**Figure 2**E). These measurements underscore the robustness of SWCNT emitters relative to conventional fluorophores and their advantage for sustained, long-term fluorescence imaging.

## F-actin–SWCNTs exhibit nonequilibrium behaviour upon myosin-induced contraction

Myosin converts the chemical free energy of ATP hydrolysis into mechanical work, generating contractile forces by binding to and sliding along actin filaments. In reconstituted F-actin networks, this motor activity produces active stresses, mesoscale remodelling, and bulk contraction, which are hallmarks of driven, nonequilibrium dynamics. To test whether our incorporated SWCNTs within filamentous actin register such motor-driven activity, we introduced myosin into the samples and monitored the SWCNT near-infrared emission alongside structural changes (visible channel) of the network.

Upon myosin addition, the F-actin network reorganized, clearly exhibiting the expected actomyosin contractility, thus confirming the transition to a nonequilibrium regime, fuelled by ATP (**Figure 3**A, **Figure S2**A, and **Movie S2**). **Figure 3**A shows the first and last dual-channel fluorescence images, along with several intermediate snapshots of a representative ROI, capturing the dynamic reorganization of the actomyosin network following myosin addition. The green channel corresponds to phalloidin-labelled F-actin, while the red channel corresponds to the NIR emission from SWCNTs embedded within the network. Before myosin introduction, the actin mesh appears homogeneously distributed, with isolated SWCNTs scattered throughout. Following the addition of myosin (t = 15 s), and over time, the network undergoes compaction with the formation of dense actin bundles (18-37 s), characteristic of nonequilibrium myosin-driven contractions, where the final frame (t = 44 s) shows a large-scale contraction of the actin network into multiple large clusters. In parallel, temporal fluctuations in the SWCNT fluorescence intensity emerge, appearing or becoming more pronounced during network contraction, indicating that the local environment of the SWCNTs responds dynamically to myosin-driven mechanical activity.

Individual SWCNT emitters within the field of view (FOV) were segmented in Fiji using the *3D Iterative Thresholding* plugin with MSER, and their NIR intensities were tracked over 50 baseline frames, followed by 100 frames after myosin addition. **Figure 3**B shows representative NIR fluorescence intensity traces, where the dashed line marks myosin addition at frame 50. Before motor activation, signals are steadier with low variance, whereas immediately afterward, the traces become intermittently bursty, with pronounced fluctuations, transient brightenings and fadings, and occasional disappear-reappear events (**Figure S2**).

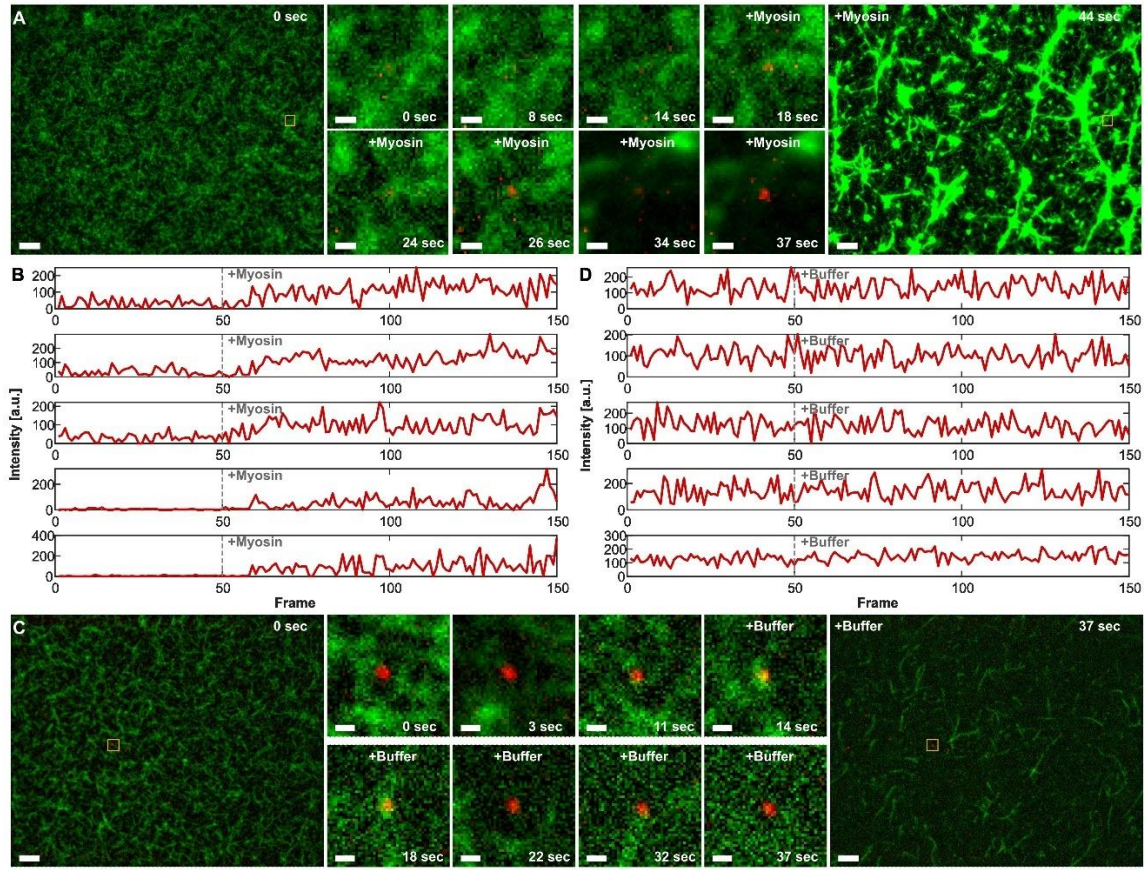

**Figure 3: Myosin-driven nonequilibrium dynamics in F-actin-SWCNT network.** A) Dual-channel representative overlay snapshots of the F-actin-SWCNT network from **Movie S2** (green: Acti-stain 488 phalloidin, $\lambda_{ex}$ = 490 nm; red: SWCNTs, $\lambda_{ex}$ = 730 nm) before and after myosin addition. Left and right panels show the full FOV at t=0 s and t= 44 s, respectively (myosin addition is at t = 15 s). Middle panels display time-lapse frames of a representative ROI (orange square in the full FOV). Scale bars: 10 μm (whole FOV) and 1 μm (ROIs). B) NIR fluorescence intensity traces of five representative ROIs. The dashed grey line marks the addition of myosin to the system. C) Overlay snapshots of the F-actin-SWCNT network from a control experiment with buffer added in place of myosin (**Movie S3**), showing the F-actin-SWCNT network before and after buffer addition, with green and red channels as in A). Left and right images correspond to the full FOV at t=0 s and t = 37 s, respectively (buffer addition is at t = 14 s). Middle panels show time-lapse frames of a representative ROI (orange square in the full FOV). Scale bars: 10 μm (whole FOV) and 1 μm (ROIs). D) NIR fluorescence intensity traces of five representative ROIs,

where buffer was added instead of myosin. The dashed grey line indicates the time of buffer addition.

To confirm that the observed fluorescence dynamics were not artifacts of fluid mixing, we performed control experiments in which an equivalent volume of buffer alone, without myosin, was added to otherwise identical F-actin-SWCNT samples (**Movie S3**). As shown in **Figure 3**C, the phalloidin-labelled F-actin network retained its homogeneous morphology throughout the time series, displaying no detectable rearrangement or contraction apart from the gradual, uniform decrease in visible fluorescence intensity expected from photobleaching.

Similarly, the SWCNT fluorescence remained consistent before and after buffer addition, without the intermittent bursting or fading behaviour characteristic of the myosin-driven traces (**Figure 3**D). The temporal profiles showed low variance and no systematic change in mean intensity following the buffer addition event (dashed line at frame 50). Together, these results rule out mechanical or optical artifacts associated with fluid addition and confirm that the pronounced intensity fluctuations observed in the presence of myosin arise specifically from active, ATP-fuelled processes within the actomyosin network.

To quantify the fluorescence intensity variability across all SWCNTs in the FOV, we computed the index of dispersion (IOD)[66], also known as the Fano factor, defined as the ratio of variance to mean. The IOD captures deviations from Poisson-like behaviour, providing a measure of temporal unpredictability or randomness in NIR emission, where larger IOD values correspond to stronger fluorescence fluctuations. For each SWCNT, IOD values were calculated using time windows of 50, 25, and 10 frames, and the data were plotted using a swarm plot overlaid by a box plot. In the myosin-treated samples (**Figure 4**A), IOD values increased markedly after myosin addition across all window sizes, indicating the onset of enhanced fluorescence fluctuations. The median IOD rose from ~10 at equilibrium to 50 post-myosin, with a transient spike approaching ~100 in the 10-frame window immediately following myosin addition, manifesting a short burst of high variability. Although the fluctuations partially relaxed over time, the post-myosin IOD values remained higher compared to pre-myosin values, reflecting greater heterogeneity in fluorescence dynamics.

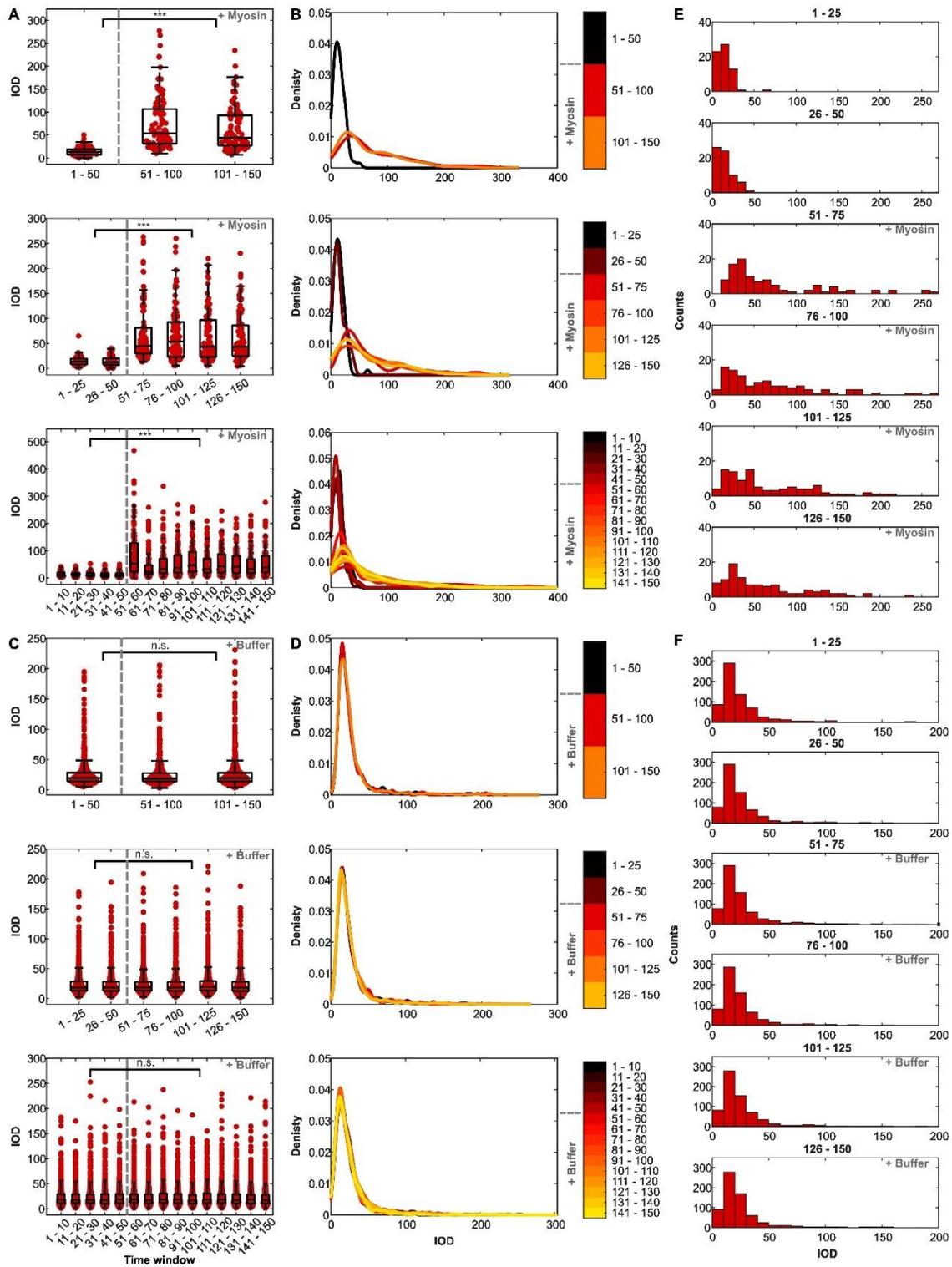

**Figure 4: F-actin-SWCNT fluorescence index of dispersion reports myosin-driven nonequilibrium dynamics.** A) Swarm plot of the distribution of the index of dispersion (IOD) for F-actin-SWCNT NIR fluorescence intensity calculated over different time-

windows (top to bottom: 50, 25, and 10 frames) from **Movie S2**, overlaid by a box plot. The dashed grey line indicates the frame at which myosin was added to the system. IOD values increased significantly after myosin addition in all window sizes (***p-value < 0.001). B) Kernel density estimation (KDE) plots showing the probability density distribution of IOD values for the same datasets and window sizes as in A). Colorbar indicates the time window of the corresponding data in terms of frame numbers. C) Swarm plot of the IOD distribution for F-actin-SWCNT NIR fluorescence intensity calculated over different time-windows (top to bottom: 50, 25, and 10 frames) from the buffer-only control **Movie S3**, overlaid by a box plot. The dashed grey line indicates the frame at which the buffer was added to the system. No significant change in IOD values was observed across all window sizes (non-significant (n.s.): p-value > 0.05). D) KDE plots showing the IOD distributions for the same datasets and window sizes as in C). E) Evolution of IOD distributions for the 25-frame analysis of **Movies S2** shown as histograms across consecutive time windows (titles indicate frame ranges), before and after myosin addition, showing a shift to higher and broader IOD values, and for F) buffer-only control (of **Movies S3**), before and after buffer addition, showing that distributions remain narrow and unchanged.

To further characterize the distributions of the IOD values, we used Kernel density estimation (KDE) analysis to estimate the corresponding probability density function. The resulting density curves clearly revealed distinct distribution profiles before and after myosin addition (**Figure 4**B) for all window sizes. The pre-myosin distributions exhibited narrow, unimodal peaks, whereas post-myosin distributions broadened and developed heavy tails, signifying the enhanced fluctuations under nonequilibrium and driven conditions.

In contrast, buffer-only control showed no significant changes. IOD values remained statistically unchanged before and after addition across all window sizes (**Figure 4**C), and the corresponding KDE plots (**Figure 4**D) preserved nearly identical distributions, with similar width and peak positions. These results confirm that the increased variance observed in the myosin-treated samples originates from active, ATP-fuelled contractile forces rather than sample perturbation.

For subsequent analysis, we focused on the 25-frame window, which offered a balance between temporal resolution and statistical robustness. Histogram representations of IOD values displayed separately for consecutive time windows further illustrate these trends. In the presence of myosin, the IOD distributions shifted toward higher and more broadly spread (**Figure 4**E), whereas in the control conditions, the distributions remained narrow and stable throughout the acquisition (**Figure 4**F).

Together, these results demonstrate that myosin-driven F-actin-SWCNTs systems can be characterized by enhanced temporal fluctuations in the NIR fluorescence, an optical signature of dynamic, nonequilibrium, ATP-driven reorganization within the actin network.

**<u>Varying Degree of Nonequilibrium Activity</u>**

To evaluate the sensitivity of the SWCNT NIR fluorescence IOD distributions to different levels of nonequilibrium activity in the actomyosin network, we repeated the experiments using reduced myosin concentrations. Lowering the myosin concentration to 50% of the original amount (see Methods section) reduced the overall contractile activity of the network, and the contraction was slower and less pronounced (**Figure S3**A-B, and **Movie S4**). Moreover, the contraction was hardly visible in the visible channel (green) due to the evident photobleaching of the phalloidin dye. Nevertheless, focusing on individual SWCNT fluorescence in the NIR channel (red), distinct fluorescence intensity modulations emerged shortly after myosin addition, even when large-scale deformation was absent in the visible channel (**Figure 5A**). Examination of individual NIR fluorescence trajectories revealed that even at 50% myosin concentration, SWCNTs exhibited clear signatures of motor-induced activity (**Figure S3**C). While most emitters showed consistent fluctuations prior to motor addition, distinct subsets displayed intermittent bursts or transient intensity changes shortly afterward. The timing and amplitude of these events varied among SWCNTs, some showing early fluctuations increase near the onset of myosin activity, while others exhibited delayed intensity modulations toward the end of the experiment, highlighting spatial heterogeneity in the local mechanical environment within the network.

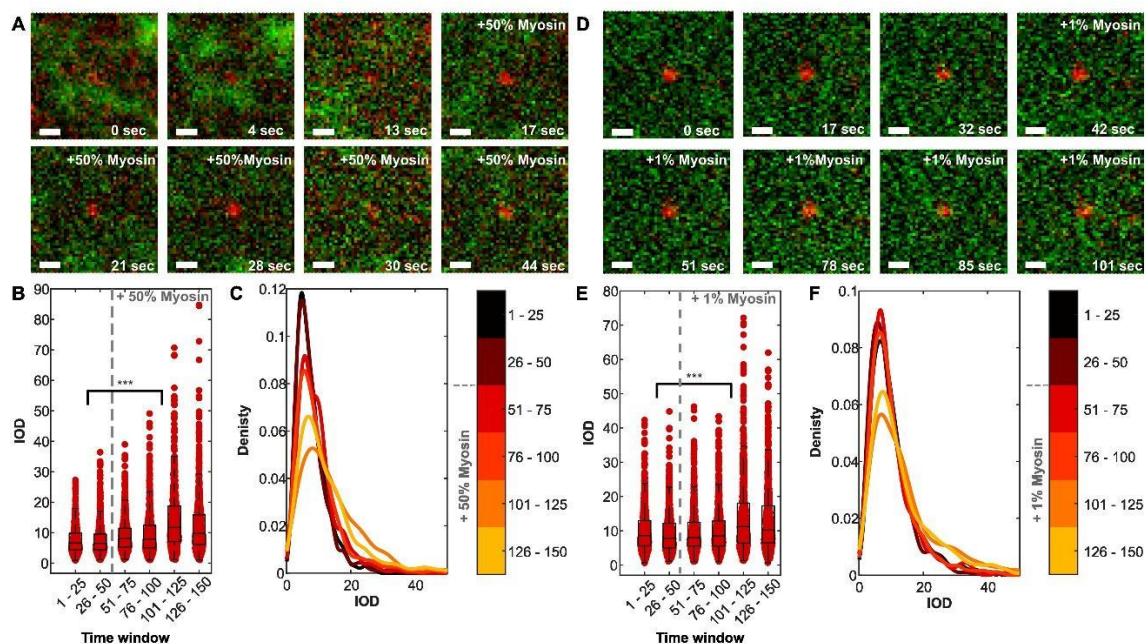

**Figure 5: Myosin-driven nonequilibrium dynamics and index of dispersion analysis in F-actin-SWCNT network with reduced myosin concentration.** A) Dual-channel representative overlay time-lapse frames of the F-actin-SWCNT network from **Movie S4** (orange squares in **Figure S3**A, B. Green: Acti-stain 488 phalloidin, $\lambda_{ex}$ = 490 nm; red: SWCNTs, $\lambda_{ex}$ = 730 nm) before and after 50% myosin addition (at t = 30 s). Scale bar: 1 µm. B) Corresponding swarm plot of the IOD distribution calculated for 25-frame windows from **Movie S4**, overlaid by a box plot. The dashed grey line indicates the frame at which 50% myosin was added to the system (***p-value < 0.001). C) KDE plots showing the probability density distribution of IOD values from the same dataset as in B). D) Dual-channel representative overlay time-lapse frames of the F-actin-SWCNT network from **Movie S5** (orange squares in **Figure S3**D,E) before and after 1% myosin addition (at t = 30 s). Scale bar: 1 µm. E) Corresponding swarm plot of the IOD distribution calculated for 25-frame windows from **Movie S5**, overlaid by a box plot. The dashed grey line indicates the frame at which 1% myosin was added to the system (***p-value < 0.001). F) KDE plots showing the probability density distribution of IOD values from the same dataset as in E).

Quantitative analysis across all emitter ROIs (~600 ROIs per movie) confirmed the response of the SWCNT fluorescence to motor activity. At 50% myosin, the IOD increased following motor addition (**Figure 5**B), though more gradually and less sharply than at full myosin concentration (**Figure 4**A). The corresponding KDE curves broadened and shifted toward higher IOD values, reflecting an overall enhancement in NIR emission intensity variability across the population (**Figure 5**C). Notably, in the final analysis window (frames 126-150), the median IOD decreased slightly, and the KDE distribution narrowed compared to the preceding window, suggesting partial relaxation of the network. However, the persistence of a long distribution tail, with some particles exhibiting even larger IOD values than before, indicates that a subset of SWCNTs continued to experience local fluctuations under the nonequilibrium conditions.

At 1% myosin, the fluorescence response was further attenuated and delayed. Although network contraction was not visible in the visible channel of the fluorescently labelled F-actin (**Figure S3**D,E, green channel), individual SWCNTs again showed emergent fluctuations after myosin addition (**Figure 5**D and **Figure S3**F). The IOD (**Figure 5**E) and KDE (**Figure 5**F) analysis revealed comparable trends of smaller but detectable increases in variability and distribution width, which appeared only in later analysis windows compared to higher-concentration experiments, consistent with slower or weaker activation of contractile events. The IOD increase became evident only at the fifth analysis window (frames 101-125), while the KDE curves showed a slightly broader distribution. Histograms of IOD values corroborated these observations, showing that for the 50% myosin condition, the distribution shifts occurred relatively quickly after myosin addition, whereas in the 1% myosin condition, it emerged gradually only in later windows (**Figure S3**G,H).

To quantitatively compare how the IOD distributions evolved under active versus control conditions, we computed the Kullback-Leibler divergence (KLD) between each post-addition distribution and the last pre-addition baseline distribution (**Figure S3**I). Specifically, for each experimental condition, we calculated

$$D_{KL}(IOD_{t_i} \parallel IOD_{t_0})$$

where $IOD_{t_i}$ denotes the probability density of IOD values within a given post-addition time window $t_i$ (e.g., frames 51-75, 76-100, 101-125, and 126-150), and $IOD_{t_0}$ corresponds to the reference distribution from the last pre-addition window (frames 26-50). The KLD between two probability distributions $p(x)$ and $q(x)$ is $D_{KL}(p \parallel q) = \int p(x) \ln(p(x)/$

$q(x)) \, dx$, and therefore quantifies the degree of distinguishability between the post-addition (of either myosin or buffer) IOD statistics and the baseline distributions.

For full-concentration myosin, the KLD increased sharply and immediately after motor addition (starting at frame 51), consistent with a rapid transition to a nonequilibrium regime. At 50% myosin, the KLD exhibited a slower, more incremental rise across successive time windows, mirroring the more gradual increase in IOD and the broadening of the distribution. At 1% myosin, divergence appeared only at later times, as the KLD remained near baseline for the first time windows after addition, and increased noticeably only in the 101-125 frame range, followed by partial relaxation in the final time window. In contrast, the buffer-only control maintained KLD values near zero with minimal variance, confirming that the IOD distributions remained stable throughout the experiment. This information-theoretic analysis therefore reinforces that the observed redistribution of fluorescence variability is a direct signature of myosin-driven, nonequilibrium dynamics.

To further characterize the transition to a nonequilibrium state, we performed stationarity analysis on the SWCNT fluorescence time series (**Figure 6**). Non-stationary signals, whose statistical properties change systematically over time, are indicative of time-irreversible dynamics[67], a characteristic of nonequilibrium systems. Each SWCNT trace was classified as nonstationary, stationary, or inconclusive, using the Kwiatkowski-Phillips-Schmidt-Shin (KPSS[68]) and Elliott-Rothenberg-Stock (DF-GLS[69]) tests (see Time Series Analysis in Methods). These statistical tools allowed us to further label the nonstationary processes as either trend-stationary (exhibiting a deterministic linear drift), unit root (possessing stochastic long-term memory), or other. To illustrate the types of temporal behaviours observed in the fluorescence data, **Figures 6A**, **6B**, **6C**, and **6D**, present representative examples of individual traces before and after the intervention, which were categorized as stationary, trend-stationarity, difference-stationarity, or inconclusive, respectively, for the different tested interventions (buffer, 1%, 50%, and 100% myosin).

Comparing the fraction of nonstationary traces across all pooled baseline conditions (before the addition of either buffer or myosin) with those in myosin-treated samples, we observed a clear dose-dependent increase (**Figure 6E**, red lines). In the absence of myosin, approximately 18% of traces were classified as nonstationary. Upon addition of myosin, this nonstationarity proportion progressively rose to ~35% of traces at 1% myosin concentration, ~73% at 50% myosin, and ~77% at full (100%) myosin concentration.

Given the stochastic nature of actomyosin network assembly, the observed stationarity fractions may vary across experimental realizations. To quantify this variability and assess the statistical significance of the dose-dependent trend, we employed a bootstrapping procedure to estimate the distribution of nonstationarity fractions. Crucially, bootstrapping relies on resampling data in fundamental units that are assumed to be independent and identically distributed (i.i.d.). In our system, however, the stationarity of neighbouring SWCNTs may be coupled. To account for this, we repeated the bootstrapping procedure across a range of models: starting from the assumption of no spatial correlation (standard bootstrap), where the stationarity classification of each single SWCNT is treated as an independent outcome, and extending to models assuming spatial correlations of varying lengths, namely Moving Block Bootstrap (MBB), where spatial blocks of neighbouring SWCNTs serve as the fundamental i.i.d. units. **Figure 6E** presents the confidence intervals (IQR and 95% CI) derived using a representative spatial block size. Despite the wide confidence intervals in the Buffer control arising from inter-experiment variability, the fraction of nonstationary traces at 1%, 50%, and 100% myosin concentrations falls distinctively outside the control interval, confirming statistical significance ($p<0.05$). This finding is robust against the assumed correlation length, as confirmed by a sensitivity analysis of the block size parameter (**Figure S4**). Beyond statistical significance, the ~4-fold increase in nonstationarity observed at high myosin concentrations (50–100%) represents a physically significant transition in the dynamical state of the actomyosin network.

Multicategory decomposition of nonstationarity across spatial block-size choices in the MBB analysis further clarifies the nature of the myosin-induced transition. Across all choices of block size, the observed increase in the fraction of nonstationary traces is primarily driven by a rise in trend-stationary traces (**Figure S5-S11**), indicating persistent, directed changes in the local environment of SWCNTs, consistent with progressive network reorganization and contractile remodelling.

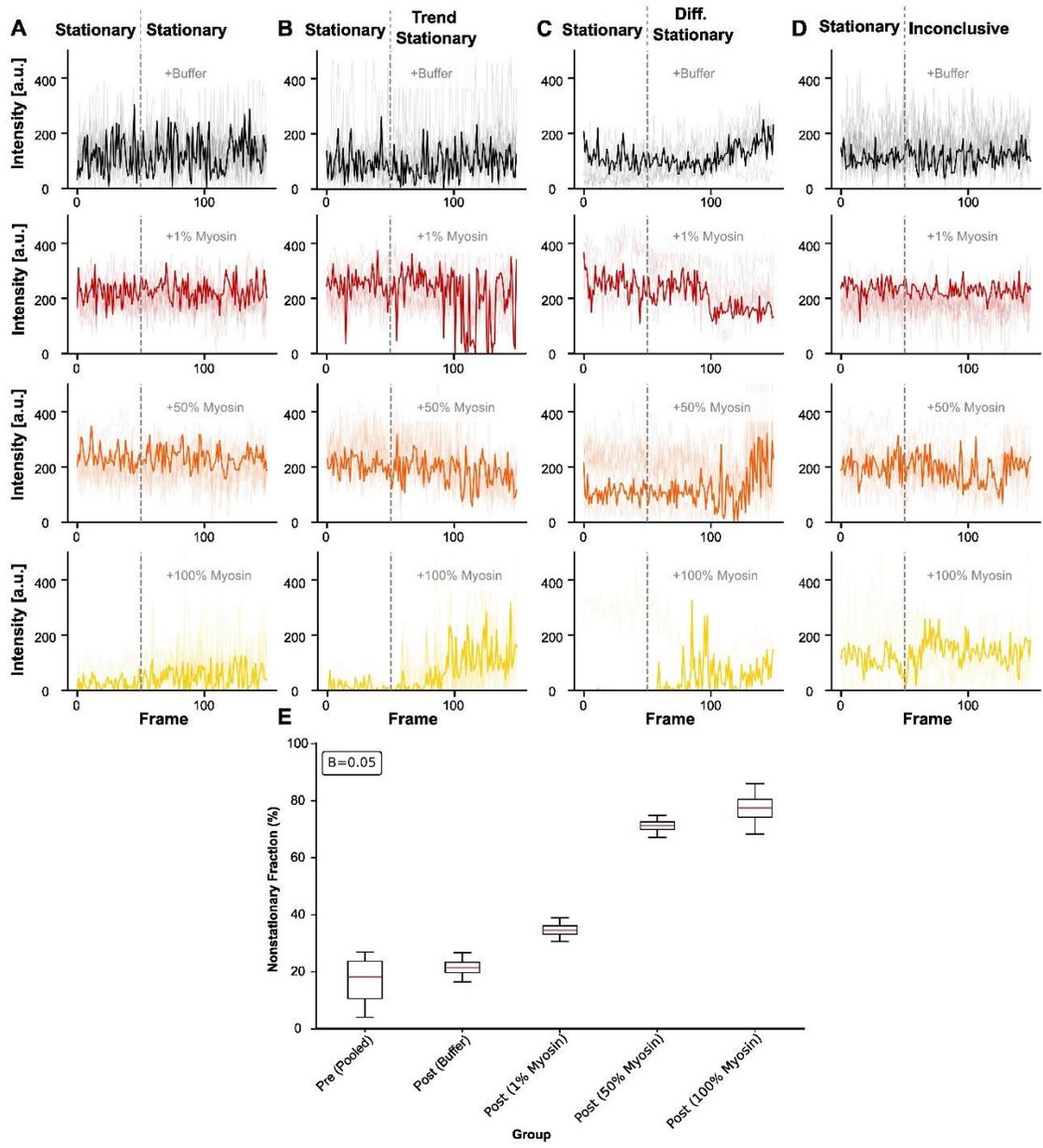

**Figure 6: Stationarity analysis of SWCNT fluorescence time series reveals a myosin-driven transition to time-irreversible dynamics.** A-D) Representative fluorescence intensity time series of individual SWCNTs. The vertical dashed line marks the intervention point (frame 50). Columns illustrate the classification of post-intervention dynamics: A) Stationary, B) Trend-Stationary, C) Difference Stationary (Unit Root), and D) Inconclusive. Rows correspond to treatment conditions: Buffer, 1% Myosin, 50% Myosin, and 100% Myosin. E) Box plots summarizing the fraction of nonstationary traces for the pooled baseline, buffer control, and myosin-treated conditions. Red lines denote the

observed fractions. Boxes denote the estimated interquartile range over repeated experiments, while whiskers extend to the 2.5th and 97.5th percentiles, representing the estimated 95% confidence interval over repeated experiments ($p=0.05$). The mean nonstationary fraction increases monotonically with myosin concentration, demonstrating a significant, dose-dependent transition to nonequilibrium dynamics compared to baseline and buffer controls.

Collectively, these results establish that SWCNTs embedded within an F-actin network are sensitive probes of active cytoskeletal actomyosin dynamics. Even minimal myosin activity, undetectable by conventional visible fluorescence imaging, produces measurable fluctuation variations in SWCNT NIR emission, allowing optical quantification of a graded nonequilibrium drive within the actin network. The NIR traces of individual SWCNTs reveal the spatiotemporal onset of motor activity, where population-level statistics (IOD and KLD between IOD distributions estimated by KDE) translate these single-emitter behaviours into system-level readouts, as distributions broaden and acquire heavy tails in response to motor addition. Further, our stationarity analysis shows that myosin addition causes a marked, dose-dependent rise in nonstationary traces, indicating drifts in the local SWCNT environment, consistent with progressive contractile remodelling. Importantly, the buffer controls verify that these signals are not artifacts of mixing. Because SWCNT emission is intrinsically non-bleaching and non-blinking, the observed variance reflects true dynamical changes in their local nanoenvironment. Practically, the sparse SWCNT loading provides spatially resolved single-particle resolution within dense protein networks while preserving the self-assembly process and motor function, enabling long-term NIR fluorescence imaging. Conceptually, the fluorescence emission of SWCNTs provides an optical readout for nonequilibrium drive, capable of detecting activity not observable in conventional visible fluorescence imaging.

## **Conclusion**

Living cells rely on continuous energy consumption to sustain nonequilibrium processes, yet experimentally resolving the onset and progression of such driven dynamics at the nanoscale remains challenging. We demonstrated that near-infrared fluorescent single-walled carbon nanotubes embedded within reconstituted actin networks act as robust optical reporters of ATP-fuelled myosin activity. By integrating SWCNTs into F-actin without perturbing network assembly, we enabled long-term, single-emitter fluorescence measurements, leveraging the intrinsic non-photobleaching SWCNT fluorescence and its sensitivity to nanoscale variations in the surrounding environment, features not achievable with conventional fluorescent dyes.

Upon the introduction of myosin motors, SWCNT NIR fluorescence intensity exhibited enhanced temporal variability, heavy-tailed distributions, and increased Kullback-Leibler divergence relative to equilibrium baselines. These changes scaled with myosin concentration and appeared even when structural contraction was not observed in the fluorescently labelled F-actin in the visible channel. Complementary stationarity analysis further revealed a substantial rise in nonstationary trajectories, indicating the emergence of irreversible, time-evolving dynamics within the network. In contrast, buffer controls showed stable and statistically stationary behaviour, confirming that the observed fluctuations arise from genuine motor-driven activity.

A central advantage of our approach is that all dynamical signatures are extracted directly from fluorescence intensity fluctuations, rather than from particle tracking, shape deformation, or spatial displacement, which are inherently constrained by diffraction-limited resolution. In contrast, the SWCNT fluorescence modulations arise from local changes in the environment surrounding each emitter, enabling the detection of motor-induced activity even when no visible structural reorganization is resolvable in the conventional fluorescence channel.

Our results demonstrate that SWCNT fluorescence intensity, independent of mechanical tracking or probe displacement, provides a powerful and minimally invasive optical modality for detecting nonequilibrium irreversibility in active biopolymer materials. This approach opens new opportunities for quantifying dissipative processes, mapping the spatial heterogeneity of motor-induced stresses, and developing nanoscale reporters for cytoskeletal mechanics and active matter.

## Methods

**SWCNT suspension:** HiPCO single-walled carbon nanotubes (SWCNTs, NanoIntegris) were dispersed using purified rabbit skeletal muscle monomeric actin (G-actin, AKL99-Cytoskeleton, Inc., Denver, CO, USA) as the dispersing agent. Briefly, 200 µg of SWCNTs were mixed with 2 mg of actin in ultrapure water to yield a final actin concentration of $10^{-3}$ wt%. The mixture was tip-sonicated (QSonica Q125, 3 mm tip, 6 W) in an ice bath for two 60-minute cycles. The resulting suspension was centrifuged at 21,000 rcf for 7 minutes to separate the individually suspended SWCNTs from aggregates and impurities[70]. Then, the supernatant was collected and the pellet discarded. Absorption spectra were recorded using an ultraviolet-visible-NIR (UV-Vis-NIR) spectrophotometer (Shimadzu UV-3600 PLUS), where sharp, distinguishable peaks indicated a successful suspension. The concentration of G-actin-SWCNT was determined spectroscopically with an extinction coefficient[71] of $\varepsilon_{632\,nm} = 0.036$ L · mg$^{-1}$ · cm$^{-1}$.

**F-actin preparation:** Filamentous actin (F-actin) network was prepared from the G-actin monomers according to the manufacturer's recommendations. First, G-actin was reconstituted to 1 mg mL$^{-1}$ in General Actin Buffer (BSA01, Cytoskeleton Inc.) supplemented with 0.2 mM ATP and 1 mM DTT. The mixture was incubated on ice for 60 minutes to depolymerize any filaments formed during freeze/thaw. For samples containing SWCNTs, 3 µL of the pre-dispersed G-actin-SWCNT suspension (at 30 mg L$^{-1}$) was added to the G-actin solution. Actin Polymerization Buffer (BSA02, Cytoskeleton Inc.) was resuspended in 100 mM Tris-HCl at pH 7.5, resulting in a 10× concentrated buffer. The buffer was added at 1:10 (v/v) to the G-actin solution (with or without the SWCNTs), and the mixture was incubated for 60 minutes at room temperature to allow polymerization. To visualize the filamentous network, the F-actin filaments were fluorescently labelled with Acti-stain 488 Fluorescent Phalloidin (PHDG1, Cytoskeleton Inc.). According to the manufacturer's recommended protocol, the F-actin-SWCNTs and the dye were diluted by a factor of 100 and 200, respectively. The absorption spectra of the F-actin-SWCNTs were recorded using the UV-Vis-NIR spectrophotometer.

**Transmission electron microscopy (TEM):** Samples included G-actin (0.3 mg mL$^{-1}$), G-actin-SWCNTs (10 mg L$^{-1}$), as well as F-actin and F-actin-SWCNTs taken directly from the polymerization mixture. A total of 30 µL of each sample was applied to a carbon-coated

TEM grid, excess liquid was removed by blotting with a filter paper, and the grids were immediately stained with 30 μL of 2 % (w/v) uranyl acetate. After a second blot to remove excess stain solution, the grids were left to dry at room temperature. The negatively stained samples were imaged on a JEM-1400Plus TEM (JEOL, Japan) operating at 80 kV. Micrographs were acquired with an SIS Megaview III camera using the iTEM imaging platform (Olympus).

**Atomic Force microscopy (AFM):** A total of 25 μl of each sample was deposited on 1 cm × 1 cm silicon wafer, allowed to dry for 1 hour at ambient conditions, rinsed twice with ultrapure water, dried under nitrogen and left at room conditions over night. Then, the wafers were placed in a desiccator for at least 2 days before the measurement. Measurements were performed using the NX12 AFM (Park Systems) with an PPP-NCHR probe (nominal frequency = 330 kHz, spring constant = 42 N m$^{-1}$) operated in non-contact mode. Images were processed in Gwyddion software.

**Near-Infrared (NIR) Fluorescence Spectroscopy:** The optical properties of SWCNTs were characterized by their NIR fluorescence emission spectra and excitation-emission maps. F-actin-SWCNTs samples were taken directly from the polymerization mixture, whereas G-actin-SWCNTs samples were diluted to 1 mg L$^{-1}$ prior to measurements. The fluorescence spectra were acquired using an inverted fluorescence microscope (Olympus IX73) coupled to a spectrograph (HRS-300SS, Princeton Instruments, Teledyne Technologies) and a liquid-nitrogen-cooled InGaAs detector (PyLoN-IR 1024-1.7, Princeton Instruments, Teledyne Technologies). A 730nm CW laser (MDL-MD-730-1.5W, Changchun New Industries) was used for excitation. The excitation-emission maps were acquired using a super-continuum white-light laser excitation (NKT-photonics, Super-K Extreme) coupled to a tuneable band-pass filter (Super-K varia, NKT-photonics, 450 nm ≤ $\lambda$ ≤ 840 nm, $\Delta\lambda$ = 2 nm) to generate wavelength-resolved excitation.

**Visible and NIR Fluorescence Imaging:** Filaments were diluted 1:100 in 100 mM Tris-HCl at pH=7.4, where the F-actin was fluorescently labelled with 70 nM Acti-stain 488 Phalloidin (PHDG1, Cytoskeleton Inc.). Aliquots of 120 μl were loaded into μ-Slide 18 Well Glass Bottom chambers (IBD-81817 #1.5H, 170±5 μm, D 263 M Schott glass, sterilized, ibidi GmbH, Germany). After acquiring 50 baseline frames, 20 μl of rabbit skeletal muscle myosin motor protein (full length, MY02, Cytoskeleton, Inc.), prepared in reaction buffer (25 mM Tris-HCl, pH 7.5, 35 mM KCl, 0.1 mM EGTA, 1 mM $MgCl_2$), was added at the indicated working concentrations (0.2 mg $mL^{-1}$, 0.1 mg $mL^{-1}$, and 2x$10^{-3}$ mg $mL^{-1}$).

Imaging was performed on an inverted fluorescence microscope (Olympus IX83) equipped with a 100× TIRF objective (UAPON100XOTIRF, 1.45 NA). Visible fluorescence (Acti-stain 488) was excited with an LED illumination system (CoolLED, pE4000) at 490 nm using a multiband dichroic mirror (Chroma, zt365-400/470/550), which also transmits the 900-1400 nm range. Fluorescence emission was detected through a multi-LED emission filter set (Chroma 89402m) and recorded on an EMCCD camera (Andor, iXon Ultra 888). SWCNT NIR fluorescence was excited with a 730 nm CW laser (MDL-MD-730-1.5W, Changchun New Industries), directed to the sample by a dichroic mirror that reflects 730 nm and transmits in the visible range (at 365, 400, 470, and 550 nm) and in the 900-1400 nm NIR range (Chroma zt730dcrb). The NIR emission of the SWCNTs was further filtered with a 900 nm long-pass filter (Chroma, ET900lp) and detected on an InGaAs camera (Raptor, Ninox 640 VIS-NIR).

**Image processing:** All images were processed by ImageJ and MATLAB. Since the EMCCD and InGaAs cameras have different pixel sizes and different pixel numbers, images were registered prior to analysis. Paired frames acquired with both cameras were rescaled, rotated, and translated to maximize their 2D autocorrelation. The resulting transformation was then applied to the full image stacks, and a common region of interest was cropped. Individual SWCNTs were segmented in Fiji using the '3D iterative thresholding' plugin with MSER as the threshold criteria. Statistical analysis of the data was performed in MATLAB using one-way ANOVA.

**Time Series Analysis:** To characterize the dynamical state of the actomyosin system from SWCNT fluorescence intensity time series, we performed stationarity analysis to classify individual signals and quantify how myosin addition affects the prevalence of nonstationary behaviour. Each time series, corresponding to a single SWCNT emitter, was divided into pre-myosin (baseline, frames 1-50) and post-myosin (frames 51-150) segments. Stationarity was assessed using two complementary statistical tests applied to each segment independently, namely, the Kwiatkowski-Phillips-Schmidt-Shin (KPSS[67]) and the Elliott-Rothenberg-Stock (DF-GLS[68]) tests. The KPSS test evaluates the null hypothesis that a series is stationary around a deterministic component, modelled as $y_t = \delta_t + \varepsilon_t$, where $\varepsilon_t$ is a stationary error and $\delta_t$ is either a constant (level-stationarity) or a linear trend $\alpha + \beta t$ (trend-stationarity). In contrast, the DF-GLS test evaluates the null hypothesis that a series contains a unit root, $y_t = y_{t-1} + u_t$, with $u_t$ stationary, implying a stochastic trend and long-term memory.

We classified each time series segment into one of four categories based on the test outcomes[68]. A series was classified as (1) Stationary if both the KPSS test failed to reject level-stationarity ($p \geq 0.05$) and the DF-GLS test rejected the unit root hypothesis ($p < 0.05$). A series was classified as (2) Trend-Stationary if it rejected level-stationarity (KPSS $p < 0.05$) but not trend-stationarity (KPSS with trend $p \geq 0.05$), and also rejected the unit root hypothesis (DF-GLS $p < 0.05$), indicating a deterministic linear drift without stochastic trend. A series was classified as (3) Unit Root if the KPSS test rejected both level- and trend-stationarity ($p < 0.05$ for both), but the DF-GLS test failed to reject the unit root hypothesis ($p \geq 0.05$), indicating a stochastic trend. Finally, (4) Inconclusive was assigned to cases that did not fit the above categories. To validate classifications, series classified as trend-stationary or unit root were transformed by linear detrending or first-differencing, respectively, and retested. The classification was retained only if the transformed series became stationary by both tests, which occurred in the vast majority of cases (see **Figure S5-S11** for fractions of inconclusive traces).

To characterize each experiment, we computed the fraction of time series in each stationarity category separately for the pre-myosin baseline phase (frames 1-50) and the post-myosin phase (frames 51-150). Our goal was to determine whether these proportions change after myosin addition and whether such changes can be attributed to the myosin intervention rather than natural inter-experiment variability.

To compute confidence intervals for these category proportions, we must account for possible spatial correlations between neighbouring SWCNTs within the 2D field of view, as standard bootstrap methods would otherwise be invalid. We therefore employed a 2D Moving Block Bootstrap (MBB), which addresses spatial dependencies by resampling square spatial blocks from the field of view rather than individual time series[72]. For each bootstrap iteration, a random set of squares, possibly overlapping spatial blocks, was selected (with replacement), and all time series within those blocks were included in the resampled data for that iteration. The fraction of each category was then computed from the resampled data, and this procedure was repeated 10,000 times to generate confidence intervals. The bootstrapping method was repeated with several block sizes to confirm the robustness of our results (see **Figure S4**).

Beyond computing confidence intervals for individual experimental phases, we needed to construct a confidence interval for the pooled baseline (0% myosin) condition across all experiments. This pooled baseline represents the natural variability in category proportions across independent experimental replicates in the absence of myosin. Since all experiments had similar field-of-view areas, we applied the same MBB procedure by sampling spatial blocks (with replacement) from the combined pool of all pre-myosin baseline data (frames 1-50 from all experiments), where each block is associated with a single replicate. This approach preserved both the spatial structure within each experiment and captured the inter-experiment variability. The resulting confidence intervals for the pooled baseline allowed us to assess whether post-myosin proportions at different myosin concentrations fell outside the natural baseline variability.

To quantify the effect of myosin addition within individual experiments, we also computed confidence intervals for the change in category proportions (post-myosin minus pre-myosin) using a paired MBB design. For each bootstrap iteration within a single experiment, the same randomly sampled set of spatial blocks was selected for both the pre-myosin (frames 1-50) and post-myosin (frames 51-150) data. This paired resampling ensured that the same spatial regions were compared before and after myosin addition, isolating the intervention effect from spatial heterogeneity. The proportion difference (post minus pre) was computed for each bootstrap iteration, and these paired proportion differences formed the resampled data from which we constructed confidence intervals for the within-experiment myosin effect.

To determine statistical significance, we compared the confidence intervals of post-myosin proportions at each concentration (1%, 50%, 100%) against the pooled baseline (0% myosin) confidence interval. A myosin concentration was deemed to have a statistically significant effect if its 95% confidence interval for the proportion of non-stationary traces did not overlap with the pooled baseline 95% confidence interval. This non-overlap criterion is conservative and ensures that observed differences exceed the natural variability present in the baseline condition. Additionally, for within-experiment paired comparisons, we assessed significance by examining whether the 95% confidence interval for the proportion difference (post minus pre) excluded zero, indicating a consistent directional change attributable to myosin addition.

Methodological Considerations: Our stationarity classification framework has inherent limitations that warrant discussion. First, the literature extensively discusses the bias towards spurious rejection of level- or trend-stationarity in favour of the unit root alternative, particularly in the case of short sample sizes[73–75]. However, our results show a rather low proportion of unit root labels, and the observed non-stationarity is primarily driven by the proportion difference between level and trend stationarity. This suggests that our findings are not artifacts of these known biases. Moreover, buffer control experiments provide a direct test of this concern: the post-buffer phase has the same extended length (100 frames) as the post-myosin phase and thus equal statistical power to detect non-stationarity, yet we observe no significant increase in non-stationary classifications for it. This demonstrates that the observed effect is specific to myosin intervention rather than an artifact of segment length or statistical power.

Second, complex dynamics such as structural breaks, regime-switching, or non-linear trends may be misclassified, as our tests assume linear models. While such dynamics could in principle occur, the fact that Inconclusive cases comprised <5% of classifications across all conditions suggests that the dominant non-stationary patterns in our data are indeed well-captured by the linear models underlying our tests. Moreover, the dose-dependent relationship between myosin concentration and non-stationarity proportion, increasing smoothly from 18% at 0% myosin to 77% at 100% myosin, provides strong evidence that we are capturing a genuine physical phenomenon rather than statistical artifacts.

Third, the Moving Block Bootstrap relies on the assumption that spatial correlations decay beyond a certain distance. We tested multiple block sizes (see **Figure S4**) and confirmed that confidence intervals remained qualitatively consistent, indicating robustness to block size choice. The conservative non-overlap criterion for significance (non-overlapping 95% confidence intervals) further reduces the risk of false positives due to bootstrap variability or residual spatial correlations.

Finally, pooling baseline data assumes that pre-myosin systems share a common stochastic process. While individual experiments may differ in initial network structure, our hierarchical bootstrap design explicitly quantifies this inter-experiment variability through the wide confidence intervals on the pooled baseline. The fact that myosin-treated conditions fall well outside these intervals indicates that the intervention effect substantially exceeds natural baseline variation.

### Acknowledgments
G.B. acknowledges the support of the Zuckerman STEM Leadership Program, the European Research Council (ERC) under the NanoNonEq project (Grant No. 101039127), the Air Force Office of Scientific Research (AFOSR) under Award No. FA9550-20-1-0426, the Army Research Office (ARO) under Grant No. W911NF-21-1-0101, the Israel Science Foundation (Grant no. 196/22), the Ministry of Science, Technology, and Space, Israel (Grant no. 1001818370), the Marian Gertner Institute for Medical Nanosystems at Tel Aviv University, the Tel Aviv University Center for AI and Data Science (TAD), the Zimin Institute for Engineering Solutions Advancing Better Lives, and the Naomi Prawer Kadar Foundation.

# Supporting Information

**Near-infrared fluorescent nanoprobes for irreversibility in nonequilibrium actomyosin networks**

Adi Hendler-Neumark[a], Itamar Magar[b], Shirel Kleiner[a], Geffen Rosenberg[a], and Gili Bisker[a,c,d,e,f],*

[a] School of Biomedical Engineering, Faculty of Engineering, Tel Aviv University, Tel Aviv 6997801, Israel

[b] School of Physics and Astronomy, Faculty of Exact Sciences, Tel Aviv University, Tel Aviv 6997801, Israel

[c] Center for Physics and Chemistry of Living Systems, Tel Aviv University, Tel Aviv 6997801, Israel

[d] Center for Nanoscience and Nanotechnology, Tel Aviv University, Tel Aviv 6997801, Israel

[e] Center for Light Matter Interaction, Tel Aviv University, Tel Aviv 6997801, Israel

[f] Sagol School of Neuroscience, Tel Aviv University, Tel Aviv 6997801, Israel

*Email: bisker@tauex.tau.ac.il

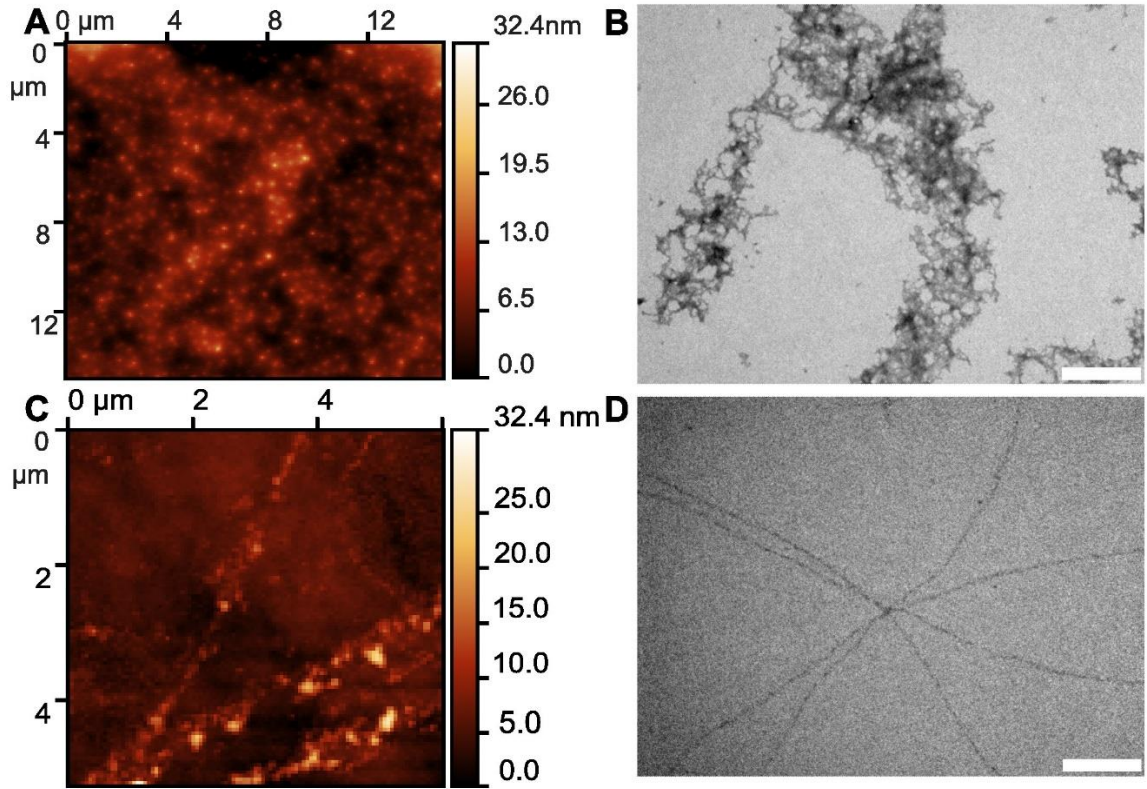

**Figure S1: Characterization of G-actin and F-actin.** A) AFM and B) TEM images of purified G-actin. Scale bar is 500 nm. C) AFM and D) TEM images of F-actin network. Scale bar is 200 nm.

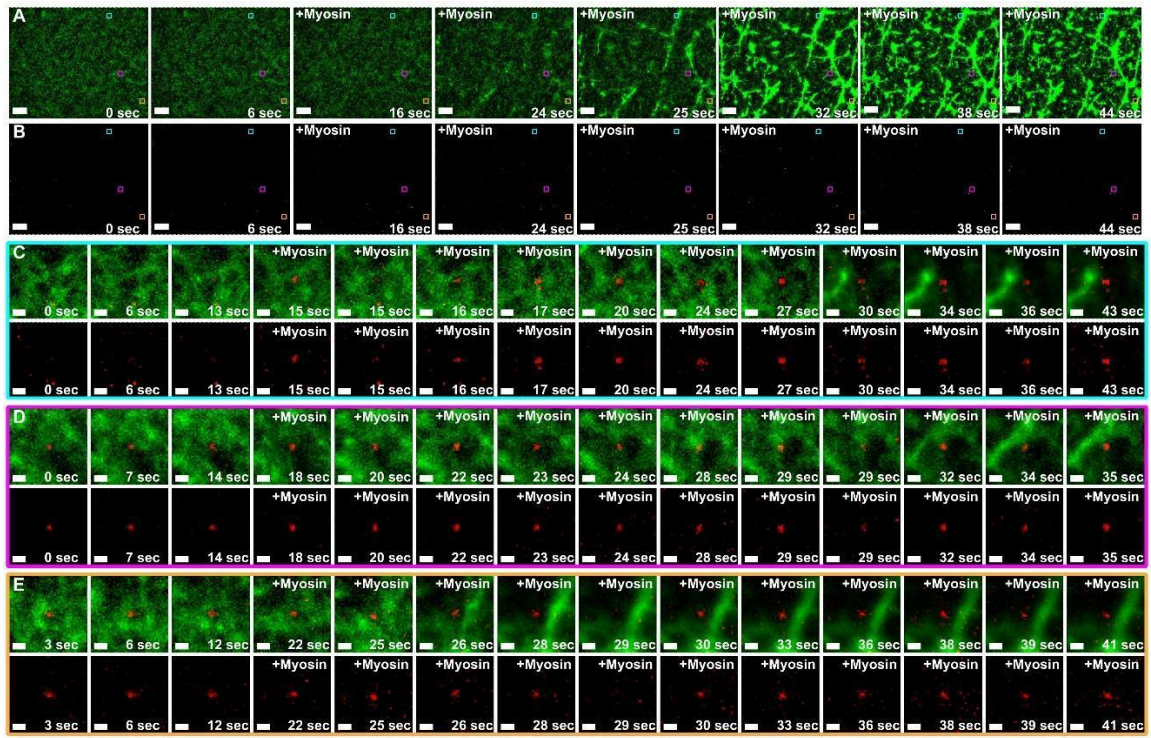

**Figure S2:** **Imaging of myosin-driven nonequilibrium dynamics in F-actin-SWCNT network.** A) Dual-channel representative overlay snapshots from **Movie S2** (green: Acti-stain 488 phalloidin, $\lambda_{ex}$ = 490 nm; red: SWCNTs, $\lambda_{ex}$ = 730 nm) showing the F-actin-SWCNT network before and after myosin addition (at t = 15 s). Scale bar: 10 μm. B) The corresponding NIR channel alone. Time-lapse frames of representative ROIs from **Movie S2** marked in C) cyan, D) magenta, and E) orange boxes, respectively. The top row in each panel is the dual-channel overlay, and the bottom row is the corresponding NIR channel alone, highlighting SWCNT fluorescence dynamics before and after myosin addition. Scale bar: 1 μm.

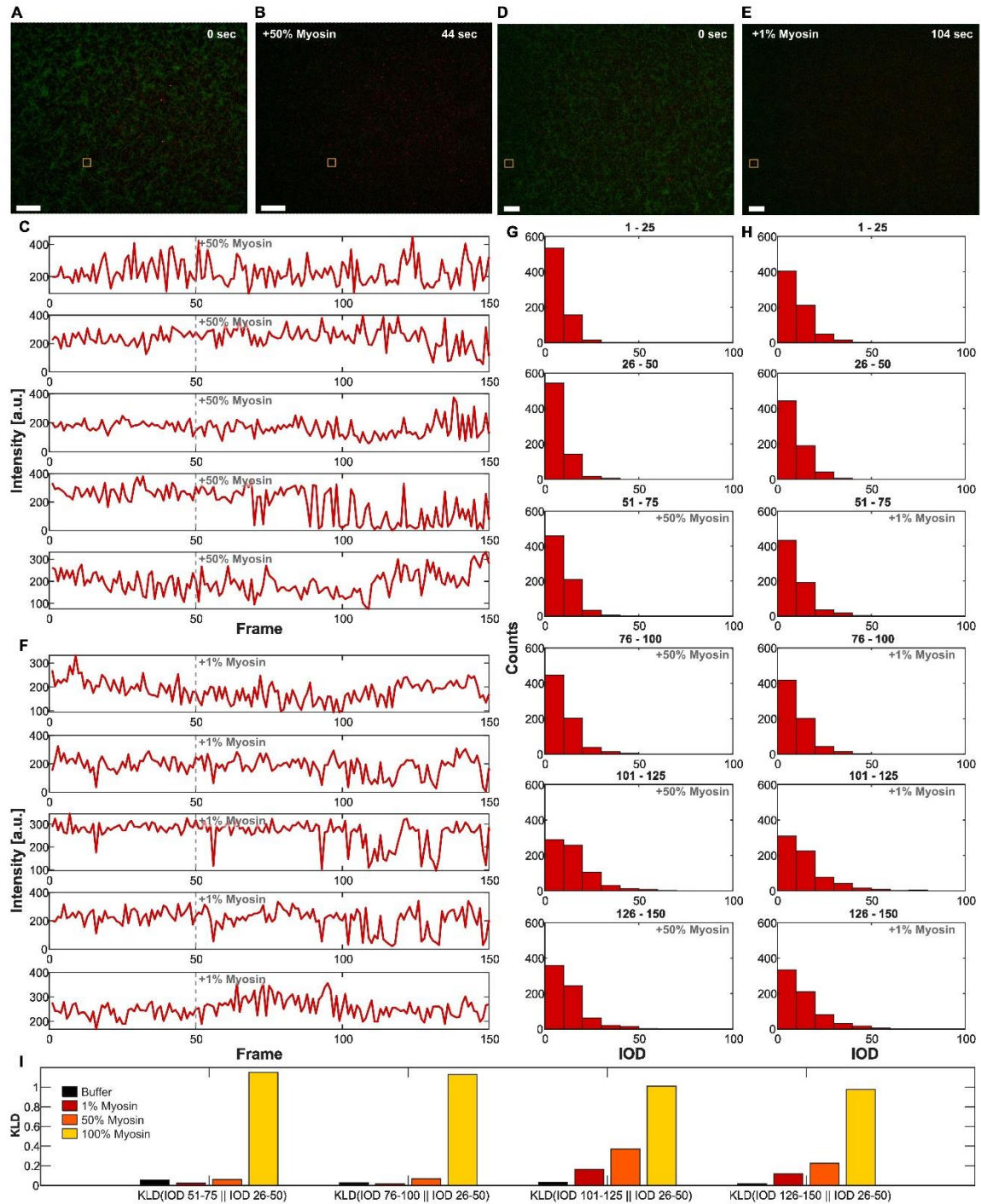

**Figure S3: Myosin-driven nonequilibrium dynamics and index of dispersion analysis in F-actin-SWCNT network with reduced myosin concentration.** A) Dual-channel representative overlay snapshot of the F-actin-SWCNT network from **Movie S4** (green: Acti-stain 488 phalloidin, $\lambda_{ex}$ = 490 nm; red: SWCNTs, $\lambda_{ex}$ = 730 nm) before and B) after 50% myosin addition. Scale bars: 10 µm. C) NIR fluorescence intensity time traces for five

representative ROIs extracted from **Movies S4** (50% myosin). The dashed grey line represents when the 50% myosin was added. D) Dual-channel representative overlay snapshot of the F-actin-SWCNT network from **Movie S5** before and E) after 1% myosin addition. Scale bars: 10 μm. F) NIR fluorescence intensity time traces for five representative ROIs extracted from **Movies S5** (1% myosin). The dashed grey line represents when the 1% myosin was added. G) Evolution of IOD distributions for the 25-frame analysis shown as histograms across consecutive time windows (titles indicate frame ranges), before and after 50% myosin addition (**Movie S4**), and H) before and after 1% myosin addition (**Movie S5**). I) KLD values between post-addition IOD distributions in various time windows and the final pre-addition baseline IOD distribution for buffer (black), 1% myosin (red), 50% myosin (orange), and 100% myosin (yellow).

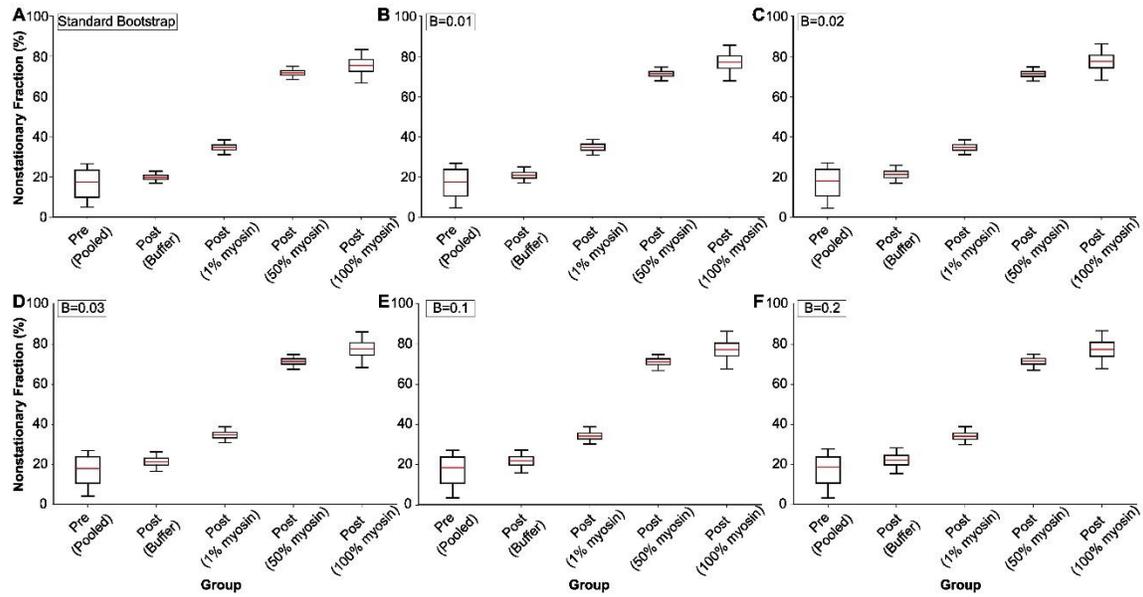

**Figure S4: Sensitivity analysis of the nonstationarity transition to spatial correlations.** To ensure robustness against spatial dependencies, we performed a systematic sensitivity analysis of the Moving Block Bootstrap (MBB) block size parameter, B, which defines the fractional area of the spatial blocks used for resampling relative to the total field of view. The fraction of nonstationary traces is plotted for pooled baseline, buffer control, and increasing myosin concentrations (1%, 50%, 100%) across a range of block sizes, representing spatial blocks covering 0% (standard bootstrap) to 20% of the total image area. Box whiskers represent 95% confidence intervals. A monotonic, dose-dependent increase in nonstationary dynamics driven by myosin remains statistically significant and qualitatively invariant across all block sizes. This confirms that the observed nonequilibrium transition is a robust physical feature of the actomyosin network, independent of the specific spatial correlation length assumed in the statistical model.

**Decomposition of stationarity categories across spatial block sizes**

The following figures (**Figures S5-S11**) present the multicategory decomposition of stationarity outcomes for different spatial block sizes (B) used in the Moving Block Bootstrap (MBB) analysis. For each block size, the upper panels illustrate the difference in stationarity category proportions (Post-intervention minus Pre-intervention) for the Buffer and Myosin-treated conditions (1%, 50%, 100%, N indicating total number of traces in each experiment), thereby isolating the intervention effect relative to the pre-intervention state of each experiment. The lower panels present the absolute proportions of stationarity categories for the individual pre-intervention phases. The transition induced by myosin is consistently characterized by a predominant increase in the Trend-Stationary component (positive difference), indicating that the network undergoes persistent, directed structural evolution.

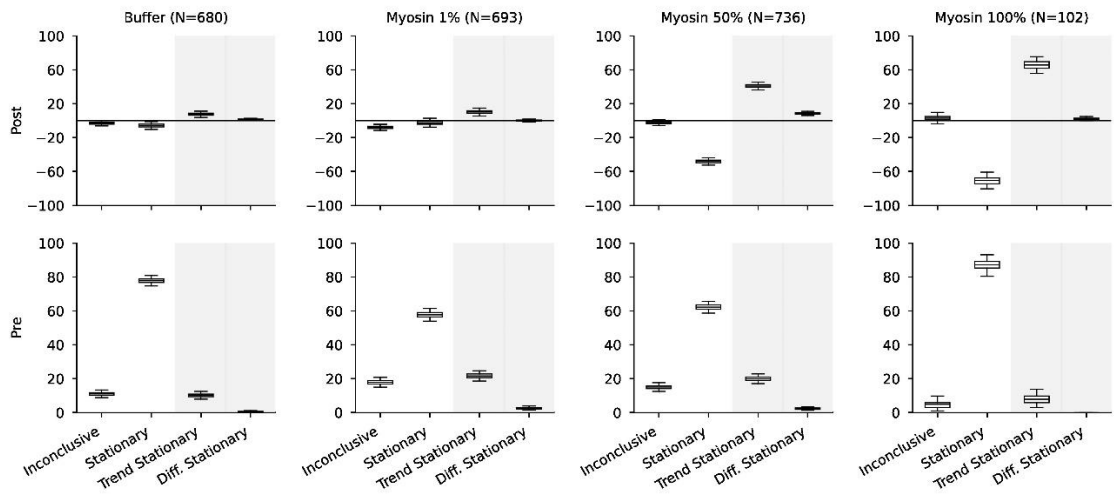

**Figure S5: Decomposition of stationarity categories using Standard Bootstrap resampling.**

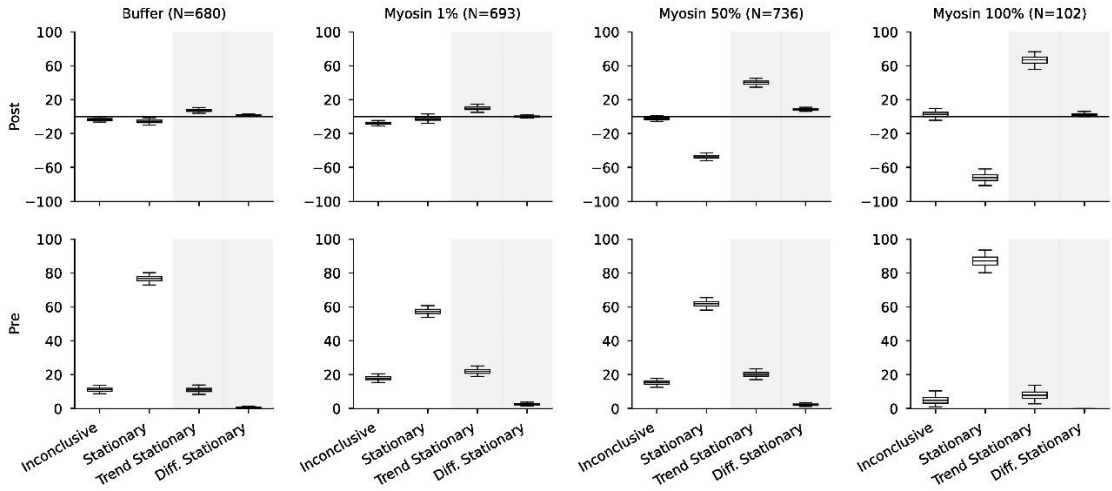

**Figure S6:** Decomposition of stationarity categories for spatial block size (B) = 0.01.

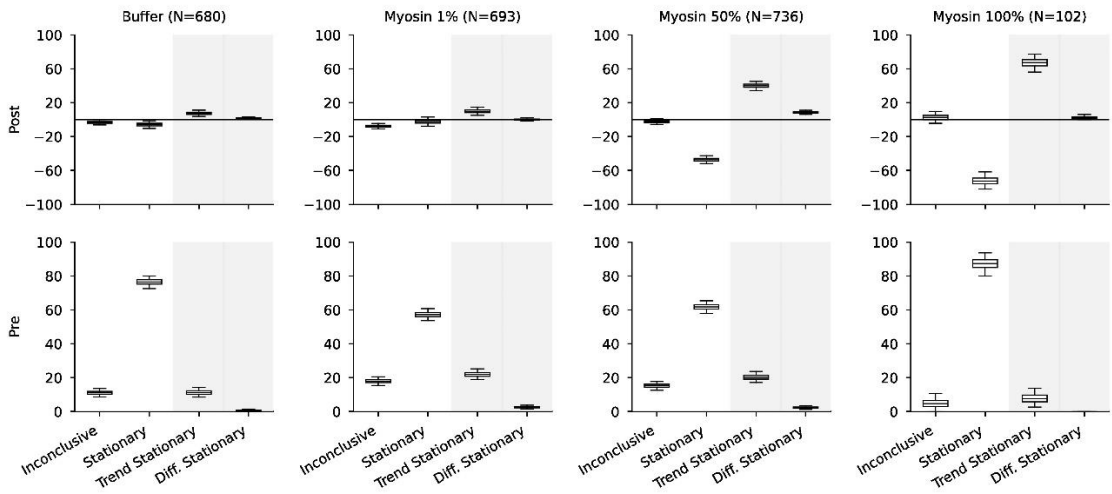

**Figure S7:** Decomposition of stationarity categories for spatial block size (B) = 0.02.

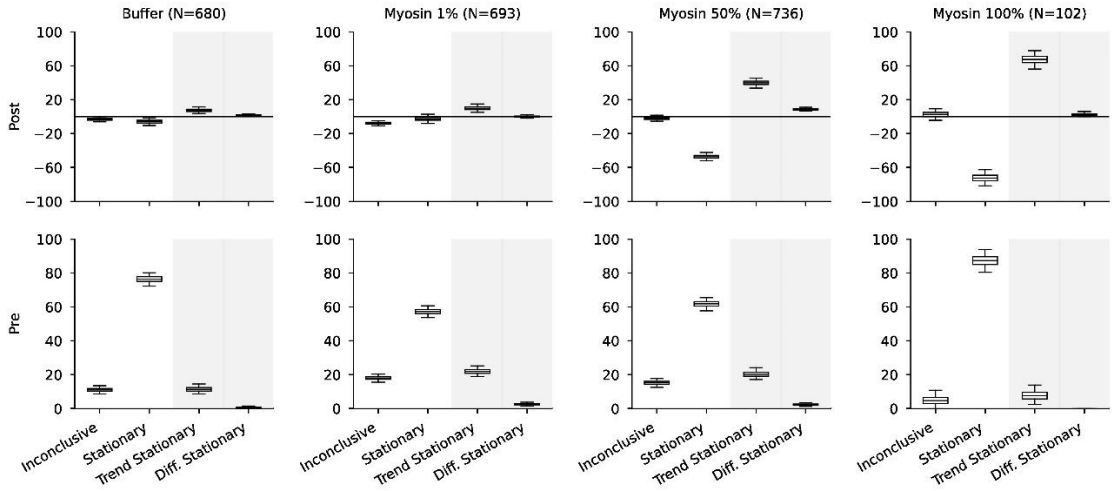

**Figure S8:** Decomposition of stationarity categories for spatial block size (B) = 0.03.

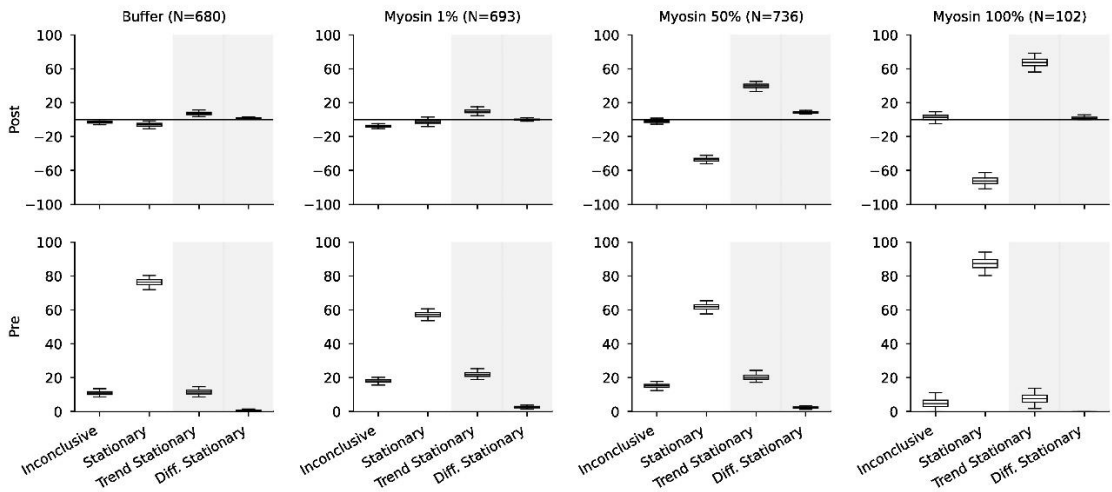

**Figure S9:** Decomposition of stationarity categories for spatial block size (B) = 0.05.

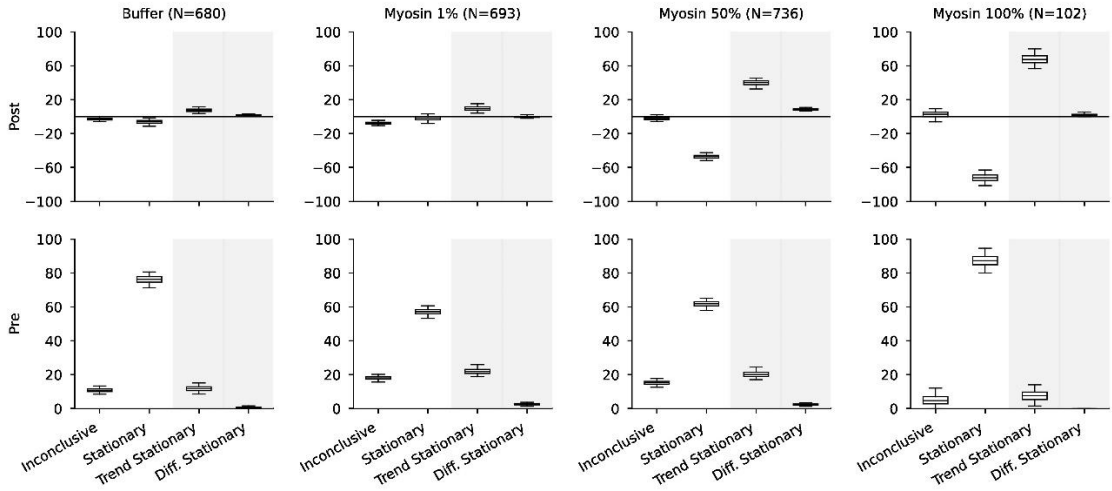

**Figure S10:** Decomposition of stationarity categories for spatial block size (B) = 0.1.

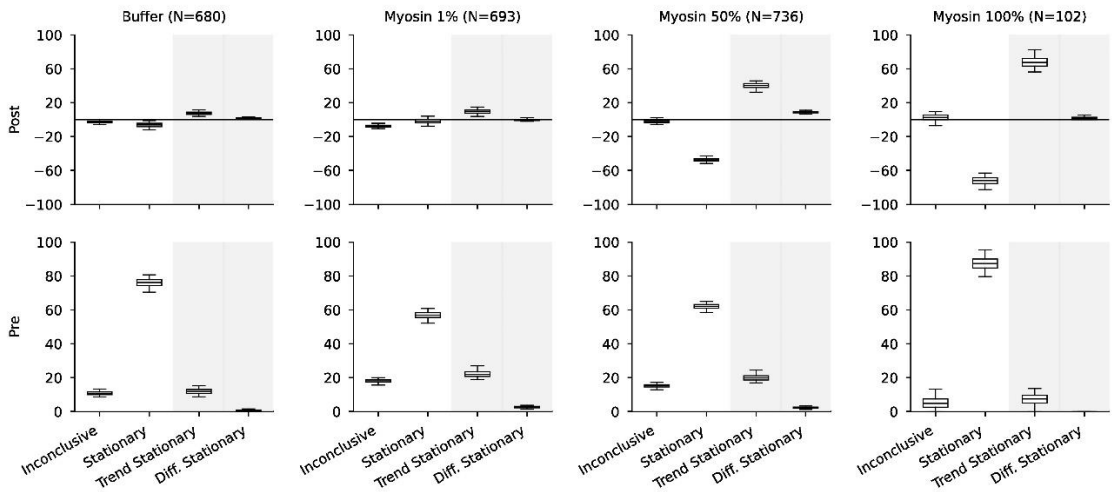

**Figure S11:** Decomposition of stationarity categories for spatial block size (B) = 0.2.

**List of Movies**

**Movie S1:** Time-lapse dual-channel overlay of F-actin-SWCNT network, showing F-actin labeled with Acti-stain 488 phalloidin (green, $\lambda_{ex}$ = 490 nm, 3 mW, 200 ms exposure), and F-actin-SWCNTs (red, $\lambda_{ex}$ = 730 nm, 680 mW, 100 ms exposure). The movie was acquired with an Olympus IX83 microscope and a 100× TIRF objective. Scale bar is 10 µm.

**Movie S2:** Time-lapse dual-channel overlay of F-actin-SWCNT network, showing F-actin labeled with Acti-stain 488 phalloidin (green, $\lambda_{ex}$ = 490 nm, 3 mW, 50 ms exposure), and F-actin-SWCNTs (red, $\lambda_{ex}$ = 730 nm, 694 mW, 200 ms exposure). Myosin was added at frame 51. The movie was acquired with an Olympus IX83 microscope and a 100× TIRF objective. Scale bar is 10 µm.

**Movie S3:** Time-lapse dual-channel overlay of F-actin-SWCNT network, showing F-actin labeled with Acti-stain 488 phalloidin (green, $\lambda_{ex}$ = 490 nm, 3 mW, 200 ms exposure), and F-actin-SWCNTs (red, $\lambda_{ex}$ = 730 nm, 442 mW, 50 ms exposure). Buffer was added at frame 51. The movie was acquired with an Olympus IX83 microscope and a 100× TIRF objective. Scale bar is 10 µm.

**Movie S4:** Time-lapse dual-channel overlay of F-actin-SWCNT network, showing F-actin labeled with Acti-stain 488 phalloidin (green, $\lambda_{ex}$ = 490 nm, 3 mW, 100 ms exposure), and F-actin-SWCNTs (red, $\lambda_{ex}$ = 730 nm, 240 mW, 200 ms exposure). At frame 51, 50% myosin was added. The movie was acquired with an Olympus IX83 microscope and a 100× TIRF objective. Scale bar is 10 µm.

**Movie S5:** Time-lapse dual-channel overlay of F-actin-SWCNT network, showing F-actin labeled with Acti-stain 488 phalloidin (green, $\lambda_{ex}$ = 490 nm, 3 mW, 100 ms exposure), and F-actin-SWCNTs (red, $\lambda_{ex}$ = 730 nm, 240 mW, 200 ms exposure). At frame 51, 1% myosin was added. The movie was acquired with an Olympus IX83 microscope and a 100× TIRF objective. Scale bar is 10 µm.